\documentclass[%
 reprint,
 amsmath,amssymb,
 aps,
prb,
]{revtex4-2}

\usepackage{graphicx}%
\usepackage{caption}
\usepackage{subcaption}
\usepackage{subcaption, calc}
\usepackage{adjustbox}
\usepackage{multirow}%
\usepackage{ragged2e}%
\usepackage{amsmath,amssymb,amsfonts}%
\usepackage{amsthm}%
\usepackage{mathrsfs}%
\usepackage{physics}%
\usepackage[per-mode=symbol]{siunitx}  
\usepackage{chemformula}  
\usepackage{upgreek}
\usepackage{orcidlink}
\usepackage[capitalise]{cleveref}
\usepackage{appendix}

\AtBeginEnvironment{appendices}{\crefalias{section}{appendix}}

\begin{document}
\title{Strain-Driven Domain Walls in Antiferromagnets}

\author{Diego De Gusem\orcidlink{0009-0008-3341-0555}}
 \altaffiliation[Dynamat ]{Department of Solid State Sciences, Ghent University, 9000 Ghent, Belgium}
\author{Arnaud Nizet\orcidlink{0009-0004-3653-2081}}
 \altaffiliation[LMGN]{Laboratory of Nanoscale Magnetic Materials and Magnonics, Institute of Materials, École Polytechnique Fédérale de Lausanne (EPFL), Lausanne 1015, Switzerland}
\author{Bartel Van Waeyenberge\orcidlink{0000-0001-7523-1661}}
\altaffiliation[Dynamat ]{Department of Solid State Sciences, Ghent University, 9000 Ghent, Belgium}


\begin{abstract}
    We derive an equation describing domain wall motion in antiferromagnets under the influence of normal strain. From this equation, we find that the domain wall moves towards positions where $\varepsilon_{xx}$ is high and $\varepsilon_{zz}$ is low. Furthermore, each strain component leads to a different terminal velocity for the same strain profile. This difference arises because both strains affect the domain wall width in opposite ways: $\varepsilon_{xx}$ reduces the width, whereas $\varepsilon_{zz}$ increases it. The model is then compared with mumax$^+$ simulations for various strain profiles, including a strain gradient, an oscillating strain, and a Rayleigh wave. The comparison shows good agreement between the analytical and numerical results. Finally, we demonstrate the potential of standing surface acoustic waves as an error correction method in racetrack memory.
\end{abstract}

\maketitle
\section{Introduction}
For technological applications, antiferromagnetic (AFM) materials offer several advantages over ferromagnetic materials \cite{advantages}. The net zero magnetization enhances stability in external fields and reduces interference with neighboring devices can allow for higher integration densities. In addition, AFM materials exhibit faster intrinsic dynamics, enabling spin textures such as domain walls and skyrmions to propagate at higher velocities than in ferromagnets \cite{fast-textures}. These properties make AFM textures promising candidates for applications such as racetrack memory \cite{racetrack, synthetic, skyrmion}, where information is stored in these textures.

Conventional approaches to drive these textures rely on applied electric currents \cite{SOT, skyrmion}, but such methods suffer from significant energy dissipation. An alternative approach is to use surface acoustic waves (SAWs), which could provide a more energy-efficient mechanism for manipulating spin textures \cite{AFM-DW-SAW, strain-DW, skyrmion-SAW}. In this work, we investigate how a spatially and temporally varying strain influences a domain wall in an AFM strip. We do this by deriving an equation of motion for the domain wall, which we then compare with micromagnetic simulations performed using mumax$^+$ \cite{mumaxplus}. Using the insights of the model, we demonstrate the ability to use two counter propagating SAWs for the purpose of racetrack memory correction.

\begin{figure}
    \centering
    \includegraphics[width=\columnwidth]{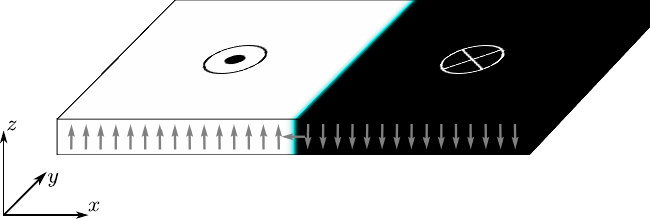}
    \caption{The Néel vector of a thin film AFM system. White and black indicate an out-of-plane and in-plane Néel vector, respectively, while blue indicates a Néel vector pointing to the left.}
    \label{fig:setup}
\end{figure}

\section{Equation of Motion}
We consider a domain wall in a thin film AFM strip, as described in \cref{fig:setup}. In the two-sublattice AFM, with sublattice magnetization $\mathbf{M}_1$ and $\mathbf{M}_2$ we define the net magnetization as $\mathbf{m} = \frac{\mathbf{m}_1 + \mathbf{m}_2}{2}$ and the Néel vector as $\mathbf{n} = \frac{\mathbf{m}_1-\mathbf{m}_2}{2}$, where $\mathbf{m}_i = \mathbf{M}_i/M_\text{s}$ and $M_\text{s}$ denotes the saturation magnetization of one sublattice. With these definitions we can assume that $\mathbf{m}^2 \ll \mathbf{n}^2 \approx 1$.

For the system in \cref{fig:setup}, the anisotropy axis is taken along the $z$-direction, and the magnetic moments are assumed to vary only along the $x$-axis. Under these conditions, the Hamiltonian of the system can be written as \cite{mumaxplus, exchange, elastics, LLG-like}.
\begin{align}\label{hamiltonian}
    \begin{split}
    \mathcal{H} = \int &\Biggl\{-\frac{8A_0}{a^2}\mathbf{m}^2 + 2A_{11}(\partial_x\mathbf{n})^2 - 2Kn_z^2\\
    &+ 2B_1\sum_{i=x,y,z}\varepsilon_{ii}n_i^2 + 2B_2\sum_{i\neq j}\varepsilon_{ij}n_in_j\Biggr\}\,\text{d}V.
    \end{split}
\end{align}
Here, $A_0$ and $A_{11}$ denote the ferromagnetic and homogeneous exchange constants, respectively, $a$ is the lattice constant, $K$ is the anisotropy constant, and $B_1$ and $B_2$ are the magnetoelastic coupling constants. In most realistic cases, the term $B_2 \sum_{i \neq j} \varepsilon_{ij} n_i n_j$ is smaller than $B_1 \sum_{i = x,y,z} \varepsilon_{ii} n_i^2$ and can therefore be neglected \cite{B1B2}.

To derive the equation of motion (EoM) we start from the LLG-like equations for antiferromagnets with low damping $\alpha$ [\citealp{LLG-like3, LLG-like2, LLG-like4}, {\citealp{LLG-like}}]
\begin{align}
    \mathbf{\dot{n}} &= (\gamma \mathbf{f_m} - \alpha\mathbf{\dot{m}}) \times \mathbf{n} \label{LLG n}\\
    \mathbf{\dot{m}} &= (\gamma\mathbf{f_n}-\alpha\mathbf{\dot{n}})\times \mathbf{n} \label{LLG m} 
\end{align}
where $\mathbf{f_m} = -\frac{1}{2\mu_0M_\text{s}}\frac{\delta E}{\delta \mathbf{m}}$ and $\mathbf{f_n} = -\frac{1}{2\mu_0M_\text{s}}\frac{\delta E}{\delta \mathbf{n}}$ are the corresponding effective fields, given by
\begin{align}
    \mathbf{f_m} &=  \frac{8A_0}{\mu_0 M_\text{s} a^2}\mathbf{m} \label{fm}\\
    \mathbf{f_n} &= \frac{2A_{11}}{\mu_0M_\text{s}}\partial_x^2\mathbf{n} + \frac{2K}{\mu_0M_\text{s}}n_z\mathbf{e}_z - \frac{2B_1}{\mu_0M_\text{s}}\sum_{x,y,z}\varepsilon_{ii}n_i\mathbf{e}_i.
\end{align}
Using \cref{fm} and (\ref{LLG n}), and taking the cross product with $\mathbf{n}$ while neglecting the term $\alpha\mathbf{\dot{m}}$, due to the low damping, we obtain an equation for $\mathbf{m}$.
\begin{align}
    \mathbf{m} &= -\frac{\mu_0M_\text{s} a^2}{8\gamma A_{0}}\mathbf{\dot{n}}\times\mathbf{n}. \label{m}
\end{align}
Taking the time derivative of this expression, we obtain
\begin{align}
    -\frac{\mu_0M_\text{s} a^2}{8\gamma A_{0}}\mathbf{\ddot{n}}\times\mathbf{n} &= \mathbf{\dot{m}}.
\end{align}
We can now use \cref{LLG m} to eliminate $\mathbf{\dot{m}}$.
\begin{align}
    -\frac{\mu_0M_\text{s} a^2}{8\gamma A_{0}}\mathbf{\ddot{n}}\times\mathbf{n} = &(\gamma\mathbf{f_n}-\alpha\mathbf{\dot{n}})\times \mathbf{n}.
\end{align}
To obtain an equation for $\mathbf{\ddot{n}}$ we again take the cross product with $\mathbf{n}$, which yields the EoM for the Néel vector.
\begin{align}
    -\frac{\mu_0M_\text{s} a^2}{8\gamma A_{0}}\mathbf{\ddot{n}} &=  \gamma\mathbf{f_n}-\alpha\mathbf{\dot{n}}\label{EOM}
\end{align}
In order to describe the dynamics of the domain wall, we introduce the collective coordinates. The
domain wall profile of the system under consideration is given in \cref{fig:setup}, where the Néel vector transitions from $+z$ on the left, to $-x$ at the DW, and to $-z$ on the right. For this Néel type domain wall, we employ the Walker ansatz is a solution, see \cref{app:walker} \cite{exchange}. The corresponding collective coordinates are then given by
\begin{align}
    &\mathbf{n}(x,t) \rightarrow \mathbf{n}_0\left(\frac{x-\chi(t)}{\Delta_0}\right)\\
    &\mathbf{n}_0 = (-\sin(\theta),0,\cos(\theta))^\intercal\\
    &\theta = 2\arctan\left(e^{\frac{x-\chi(t)}{\Delta_0}}\right).
\end{align}
Here, we define $\chi(t)$ as the domain wall position and $\Delta_0 = \sqrt{\frac{A_{11}}{K_\text{eff}}}$, with $K_\text{eff} = K + B_1(\varepsilon_{xx} - \varepsilon_{zz})$, as the domain wall width. From this equation, we can already see that a $\varepsilon_{xx}$ ($\varepsilon_{zz}$) strain increases (decreases) the domain wall width. Also notice that the domain wall width becomes time-dependent when the strain is time-dependent. However, it has been reported that AFM domain walls move at a speeds $v$ close to the magnon group velocity $c=\frac{\gamma}{\mu_0M_\text{s}}\sqrt{-32\frac{A_{11}A_{0}}{a^2}}$, we must account for the Lorentz contraction of the domain wall width $\Delta_0\rightarrow\Delta=\Delta_0\sqrt{1-v^2/c^2}$ \cite{relativistic, LLG-like}. This changes the collective coordinates and ansatz to
\begin{align}
    &\mathbf{n}(x,t) \rightarrow \mathbf{n}_0\left(\frac{x-\chi(t)}{\Delta}\right)\\
    &\mathbf{n}_0 = (-\sin(\theta),0,\cos(\theta))^\intercal \label{n0}\\
    &\theta = 2\arctan\left(e^{\frac{x-\chi(t)}{\Delta}}\right), \label{theta}
\end{align}
To obtain the EoM for the DW, we multiply \cref{EOM} by $\partial_x \mathbf{n}$ and integrate over the volume of the magnet, while only keeping terms up to first order in strain as described in \cref{app:math}. This results in the EoM for the domain wall position $\chi$
\begin{align}
    \frac{m_{\text{eff}}}{1-\frac{\dot{\chi}^2}{c^2}}\ddot{\chi} &= F - \left(\Gamma_\alpha + \Gamma_\Delta\right)\dot{\chi}. \label{EOM DW}
\end{align}
In this equation, we defined the effective mass, the force, and the damping parameters as
\begin{align}
    m_{\text{eff}} &= -\frac{\mu_0^2M_s^2 a^2}{8\gamma^2 A_0 \Delta}t_yt_z\\
    F &= -B_1\Delta\frac{\text{d}(\varepsilon_{xx}-\varepsilon_{zz})}{\text{d}x}t_yt_z\\
    \Gamma_\alpha &= \frac{\alpha\mu_0M_\text{s}}{\gamma\Delta}t_yt_z\\
    \Gamma_\Delta &= m_{\text{eff}}\frac{B_1(\dot{\varepsilon}_{xx} - \dot{\varepsilon}_{zz})}{2K_{\text{eff}}},
\end{align}
where the subscripts $\alpha$ and $\Delta$ denote that the damping term originates from the Gilbert damping and the domain wall width, respectively. From these equations, we can already observe that the $xx$-strain and $zz$-strain have opposite effects on the domain wall. In particular, under a $xx$($zz$)-strain, the domain wall moves toward regions of higher (lower) strain. We also note that $\Gamma_\Delta$ depends on the time variation of the strain, and therefore vanishes for time-independent strain profiles. We can further determine the terminal velocity by using $\ddot{\chi}=\SI{0}{\meter\per\second^2}$.
\begin{align}
    \dot{\chi} &= -\frac{F}{\Gamma_\alpha + \Gamma_\Delta}\\
    &= \frac{-\gamma B_1\Delta^2}{\alpha\mu_0M_\text{s} - \frac{\mu_0^2M_s^2 a^2}{16\gamma A_0}\frac{B_1(\dot{\varepsilon}_{xx} - \dot{\varepsilon}_{zz})}{K_{\text{eff}}}}\frac{\text{d}(\varepsilon_{xx}-\varepsilon_{zz})}{\text{d}x}.\label{term}
\end{align}

\section{Verification of the EoM}\label{sec:valid}
We now compare the predictions of the EoM  with micromagnetic simulations performed in mumax$^+$ for the domain wall dynamics resulting for several strain profiles. 

In these simulations, we use material parameters that closely resemble those of the antiferromagnet \ch{NiO} \cite{NiO-exchange,NiO-damping,AFM-strain-DW-motion,NiO-elastic}. We take a saturation magnetization $M_{\text{s}}$ of \SI{425}{\kilo\ampere\per\meter}. The ferromagnetic exchange constant $A_{11}$ and the homogeneous exchange constant $A_0$ are set to \SI{4.4}{\pico\joule\per\meter} and \SI{-46}{\pico\joule\per\meter}, respectively. The lattice constant is chosen as $a = \SI{0.42}{\nano\meter}$. 
The uniaxial anisotropy $K$ is oriented along the $z$-axis and has a value of \SI{85.7}{\kilo\joule\per\meter^3}, while the Gilbert damping is taken as $\alpha = \num{2.1d-4}$. The magnetoelastic coupling parameters $B_1$ and $B_2$ are \SI{-15}{\mega\joule\per\meter^3} and \SI{-8.5}{\mega\joule\per\meter^3}, respectively.

The simulation consists of cells of size \qtyproduct{1 x 1 x 1}{\nano\meter}, forming a \numproduct{4096 x 1 x 1} grid, with periodic boundary conditions applied along the $y$-direction. In mumax$^+$ the domain wall position is determined by interpolating where the $z$-component of the Néel vector is equal to zero. For the model we integrate \cref{EOM DW} numerically using the RK4 method \cite{RK4}.

\subsection{Static Strain Gradient}\label{sec:gradient}
\begin{figure}
    \centering
    \includegraphics[width=\columnwidth]{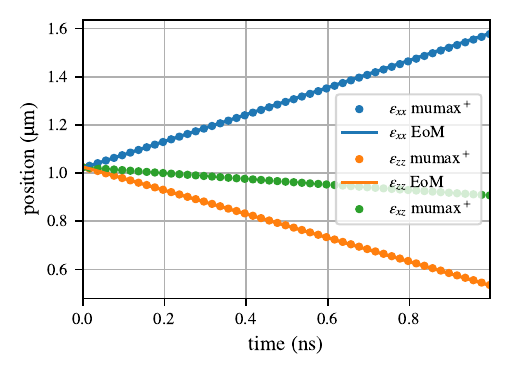}
    \caption{AFM domain wall position as a function of time under an applied gradient of $\SI{0.14}{\%}/\SI{4.096}{\micro\meter}$. The dots represent data points obtained from the mumax$^+$ simulation, while the lines show the result of integrating \cref{EOM DW}. Blue, orange and green data corresponds to the simulation with only $\varepsilon_{xx}$ strain, $\varepsilon_{zz}$ strain and $\varepsilon_{xz}$ strain, respectively.}
    \label{fig:gradient_pos}
\end{figure}

We apply a static strain gradient that increases from \SI{0}{ppm} at $x=\SI{0}{\micro\meter}$ to \SI{0.14}{\%} at $x=\SI{4.096}{\micro\meter}$. We already know that $\Gamma_\Delta$ is irrelevant for a static gradient. The results of these simulations are shown in \cref{fig:gradient_pos} . As expected, $\varepsilon_{xx}$ and $\varepsilon_{zz}$ strains have  opposite effects on the domain wall motion, while the effect of shear strain is low. The difference between the simulation and the model is only about \SI{3}{\nano\meter} after \SI{1}{\nano\second}, corresponding to three simulation cells, while the domain wall has already traveled approximately about \SI{500}{\nano\meter}. This corresponds to a deviation of about \SI{0.6}{\percent}, demonstrating good agreement between the model and the simulation.

\begin{figure}
    \centering
    \includegraphics[width=\columnwidth]{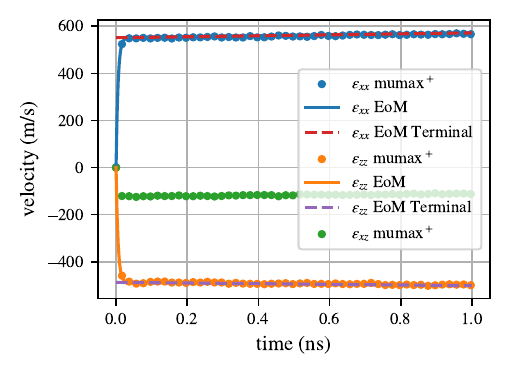}
    \caption{The domain wall velocity as a function of time under an applied strain gradient of $\SI{0.14}{\%}/\SI{4.096}{\micro\meter}$. The dots represent data obtained from the mumax$^+$ simulation, while the lines are the result of integrating \cref{EOM DW}. The dashed lines indicate the terminal velocity obtained from \cref{term}. Blue, orange and green data corresponds to the simulation with only $\varepsilon_{xx}$ strain, $\varepsilon_{zz}$ strain and $\varepsilon_{xz}$ strain, respectively.}
    \label{fig:gradient_vel}
\end{figure}

We further analyze the velocity in function of time in \cref{fig:gradient_vel} and observe that the magnitude of the terminal domain wall velocity under $\varepsilon_{xx}$ strain is \SI{200}{\meter\per\second} higher than under $\varepsilon_{zz}$ strain. This can be understood by examining the domain wall width in the terminal velocity in \cref{term}. In these equations we see that the terminal velocity linearly depends on $\Delta$ and that $xx$-strain increases $\Delta_0$ while $zz$-strain decreases it, leading to higher and lower terminal velocities, respectively. We can also see that the absolute terminal velocity of the domain wall influenced by shear strain is only about \SI{100}{\meter\per\second}, which is only a sixth (fifth) of a domain wall under influence of $\varepsilon_{xx}$ ($\varepsilon_{zz}$) strain.

\subsection{Strain Wave}
Here we apply a sinusoidal strain of the form $\varepsilon_0\sin(kx-2\pi ft)$, where $\varepsilon_0=\SI{0.14}{\%}$ is the amplitude, $k=\frac{2\pi}{\SI{250}{\nano\meter}}$ is the wavenumber, and $f=\SI{10}{\giga\hertz}$ is the frequency. In this case, the strain is time-dependent, meaning that all terms in \cref{EOM DW} contribute.

\begin{figure}
    \centering
    \includegraphics[width=\columnwidth]{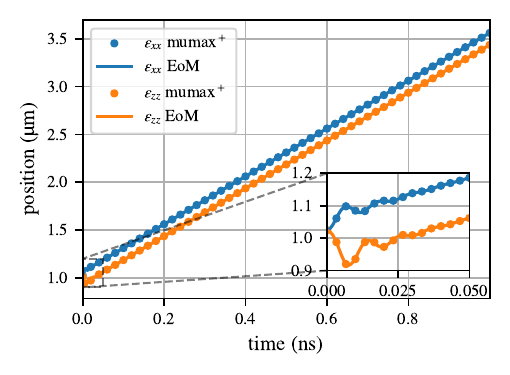}
    \caption{AFM domain wall position as a function of time under an applied sinusoidal strain of the form $\varepsilon_0\sin(kx-2\pi ft)$. The amplitude $\varepsilon_0$, frequency $f$ and wavenumber $k$ are \SI{0.14}{\%}, \SI{10}{\giga\hertz} and $\frac{2\pi}{\SI{250}{\nano\meter}}$, respectively. The dots represent data obtained from mumax$^+$ simulations, while the lines are the result of integrating \cref{EOM DW}. Blue data correspond to $\varepsilon_{xx}$ strain, wile the orange data correspond to $\varepsilon_{zz}$ strain. The inset shows a zoom on the data between \SI{0}{\nano\second} and \SI{0.05}{\nano\second}, highlighting the oscillatory behavior of the domain wall.}
    \label{fig:osc_pos}
\end{figure}

When examining the domain wall position in \cref{fig:osc_pos}, we observe that both the $\varepsilon_{xx}$ and $\varepsilon_{zz}$ case ultimately move the domain wall in the same direction. This can be explained by the domain wall becoming trapped in a local extremum (a maximum for $\varepsilon_{xx}$ and a minimum for $\varepsilon_{zz}$) and subsequently being dragged along by the strain wave. This capture process is visible in the inset of \cref{fig:osc_pos}, where the domain wall oscillates around an extremum.

\begin{figure}
    \centering
    \includegraphics[width=\columnwidth]{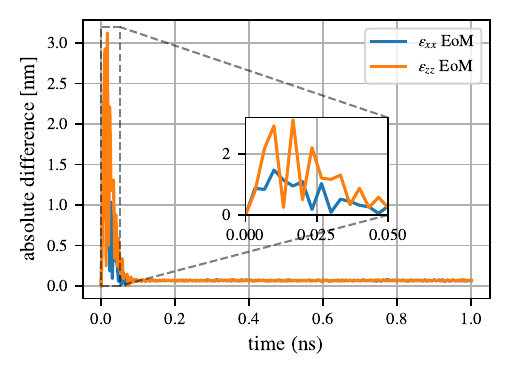}
    \caption{Absolute deviation of the domain wall position from the model compared to mumax$^+$ as a function of time under an applied sinusoidal strain. The blue line corresponds to the difference in the $\varepsilon_{xx}$ strain simulation, while the orange line indicates the difference of the $\varepsilon_{zz}$ strain simulation. The inset is a zoom on the data between \SI{0}{\nano\second} and \SI{0.05}{\nano\second}, highlighting the differences more clearly.}
    \label{fig:osc_err}
\end{figure}

From the analysis of \cref{fig:osc_err}, we observe that the deviation between mumax$^+$ and the EoM initially increases to approximately \num{3} cells, which is attributed to the rapid oscillations of the domain wall around the extrema of the sinusoidal profile. Importantly, this difference is comparable to that in \cref{sec:gradient}, and once the domain wall has settled in an extrema of the strain, the difference decreases to below one cellsize and remains constant.

\subsection{Rayleigh Waves}
A Rayleigh wave is a type of surface acoustic wave that is of particular interest in magnetoelastic systems, as it can be efficiently excited using interdigital transducers on a piezoelectric substrate \cite{SAW}. In the following, we consider a Rayleigh wave propagating along the $x$-direction, which can be described by the following strain components \cite{Rayleigh}
\begin{align}
\begin{bmatrix}
\varepsilon_{xx}\\
\varepsilon_{yy}\\
\varepsilon_{zz}\\
\varepsilon_{xy}\\
\varepsilon_{xz}\\
\varepsilon_{yz}\\
\end{bmatrix} &= \begin{bmatrix}
    \varepsilon_0\sin(kx - \omega t)\\
    0\\
    -\varepsilon_0\sin(kx - \omega t)\\
    0\\
    \varepsilon_0'\cos(kx - \omega t)\\
    0
\end{bmatrix}.
\end{align}
Note that these expressions also include shear strain, which we neglected in the derivation of the equation of motion. However, the shear component is smaller than the normal components. Also note that the $\varepsilon_{xx}$ and $\varepsilon_{zz}$ components have equal magnitude but opposite sign, implying that they have the same effect on the domain wall motion.

For the simulations, we use the amplitudes $\varepsilon_0$ and $\varepsilon_0'$ of \SI{0.14}{\%} and \SI{0.0525}{\%}, respectively, and a frequency $f$ and wavenumber $k$ of \SI{10}{\giga\hertz} and $\frac{2\pi}{\SI{250}{\nano\meter}}$.

\begin{figure}
    \centering
    \includegraphics[width=\columnwidth]{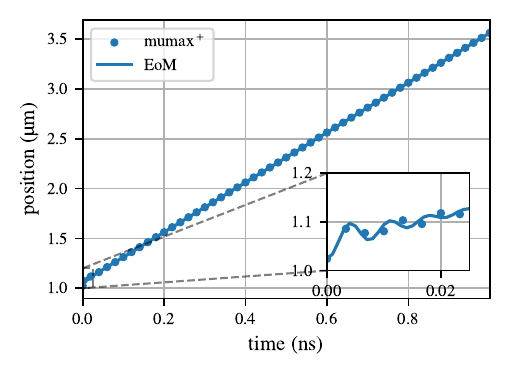}
    \caption{AFM domain wall position as a function of time under an applied Rayleigh wave. The frequency $f$, wavenumber $k$, and amplitudes $\varepsilon_0$ and $\varepsilon_0'$ are \SI{10}{\giga\hertz}, $\frac{2\pi}{\SI{250}{\nano\meter}}$, \SI{0.14}{\%}, and \SI{0.0525}{\%}, respectively. The dots represent data obtained from the mumax$^+$ simulation, while the line shows the result of integrating \cref{EOM DW}. The inset shows a zoom of the data between \SI{0}{\nano\second} and \SI{0.02}{\nano\second}, highlighting the oscillatory behavior of the domain wall and the effect of the shear strain.}
    \label{fig:saw_pos}
\end{figure}

\begin{figure}
    \centering
    \includegraphics[width=\columnwidth]{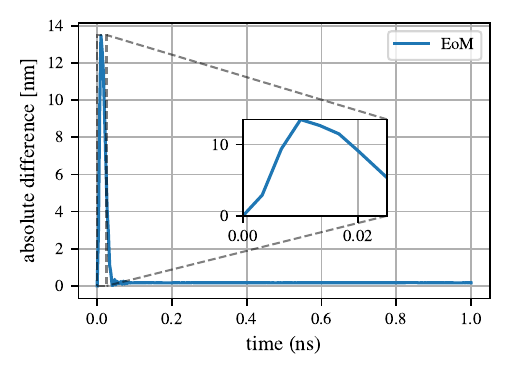}
    \caption{Absolute deviation of the domain wall position from the model compared to mumax$^+$ as a function of time under an applied Rayleigh wave. The inset shows a zoom on the data between \SI{0}{\nano\second} and \SI{0.02}{\nano\second}, highlighting the differences due to the shear strain more clearly.}
    \label{fig:saw_err}
\end{figure}

From the results shown in \cref{fig:saw_pos} and \cref{fig:saw_err}, we observe that the overall trend of the model agrees well with the simulation. However, a larger initial deviation is present due to the influence of the shear strain. The $\varepsilon_{xz}$ component tends to force the domain wall to localize at position where this component vanishes (as described in \cref{app:shear}), which coincides with the extrema of the norm strains. This introduces an additional force on the domain wall that is not captured in the model.

\section{Racetrack Error Correction}
A magnetic racetrack is a device in which magnetic domains store information, such as bits \cite{racetrack}. The domain walls are pinned at notches and can be driven from one notch to the next using an external driving force, allowing the domains to be transported toward a readout element. However, reliable operation is not always guaranteed, as a domain wall may fail to reach the intended notch and become trapped between two notches. Besides driving a domain wall, surface acoustic waves (Rayleigh waves) can be used to mitigate this issue. From the previous sections, we learned that a DW prefers regions where the $\varepsilon_{xx}$ strain is high and the $\varepsilon_{zz}$ strain is low. By applying two identical counter-propagating SAWs, a standing wave can be generated in which the antinodes coincide with the positions of the notches. This standing wave then drives the DW away from the nodes (between the notches) and toward the antinodes, where it can be captured by a notch.

We simulate this mechanism in mumax$^+$ again using the \ch{NiO} parameters described in \cref{sec:valid}. The system is defined on a grid with cell size \qtyproduct{1 x 1 x 1}{\nano\meter}, consisting of \numproduct{1024 x 512 x 1} cells. Three equally spaced notches are included, modeled as triangular indentations in the simulation lattice with a base of \SI{20}{\nano\meter} and a height of \SI{15}{\nano\meter}.

\begin{figure}
    \centering
    \includegraphics[width=\columnwidth]{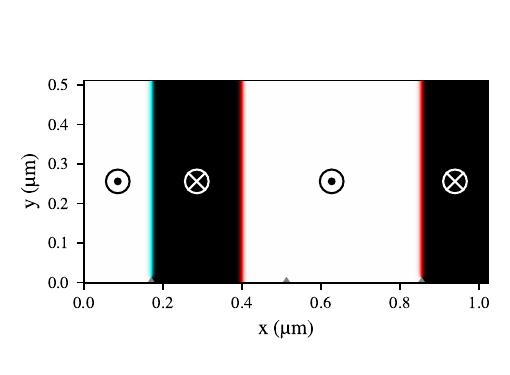}
    \caption{The Néel vector in the initial state of the magnet, containing three domain walls and three equally spaced pinning sites. White and black indicate an out-of-plane and in-plane Néel vector, respectively, while blue and red indicate a Néel vector pointing to the left and right, respectively. The first and last domain walls are pinned on the pinning sites, while the second is unpinned at the left of the middle pinning site at \SI{0.4}{\micro\meter}.}
    \label{fig:racetrac_init}
\end{figure}

The initial magnetization configuration contains three domain walls: one located at the first notch, one positioned to the left of the second notch at \SI{400}{\nano\meter}, and one at the third notch. This configuration is shown in \cref{fig:racetrac_init}. For this setup, the SAW wavelength is chosen such that the extrema of $\varepsilon_{xx}$ and $\varepsilon_{zz}$ coincide with the three notches, which corresponds to a wavelength of $\frac{1024}{6}$\unit{\nano\meter}. The frequency is fixed at \SI{10}{\giga\hertz}, and the SAW amplitudes are \SI{300}{ppm} for $\varepsilon_{xx}$ and $-\varepsilon_{zz}$, and \SI{112.5}{ppm} for $\varepsilon_{xz}$.

\begin{figure}
    \centering
    \includegraphics[width=\columnwidth]{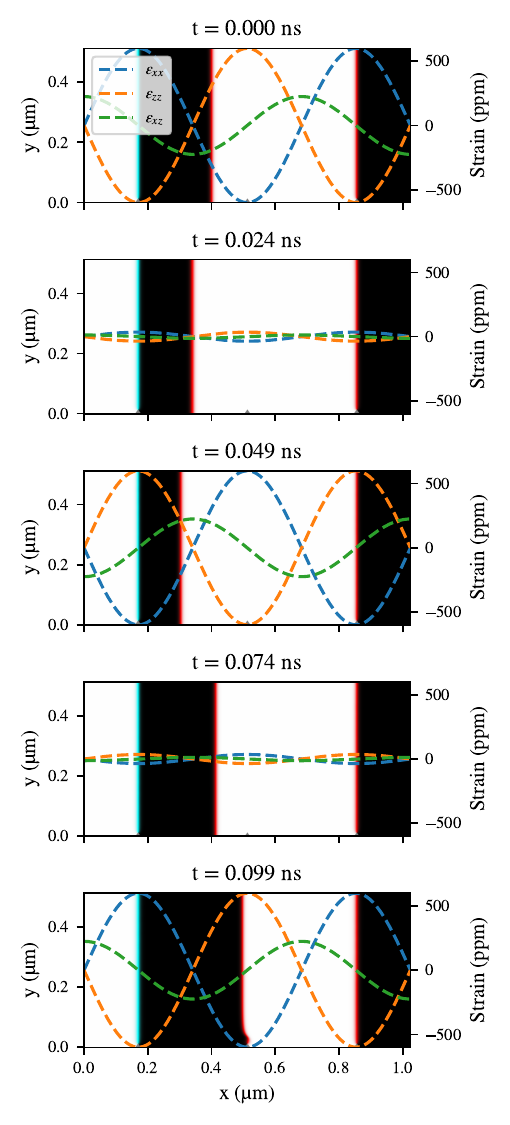}
    \caption{The Néel vector and strain components at different time points in the simulation. The strains $\varepsilon_{xx}$, $\varepsilon_{zz}$ and $\varepsilon_{xz}$ oscillate, representing a standing wave generated by two counter propagating surface acoustic waves. The images show that the central domain wall is affected by the strains, while the pinned domain walls remain unaffected. As a result of this motion, the central domain wall becomes pinned at the central pinning site after \SI{0.099}{\nano\second}.}
    \label{fig:racetrac_shots}
\end{figure}

Snapshots of the simulation are shown in \cref{fig:racetrac_shots}. At \SI{0}{\nano\second}, we predict that the domain wall will propagate toward the left due to the applied strains. This can be observed in the second frame, where the domain wall has indeed moved to the left. However, due to the oscillation of the standing wave, the domain wall now experiences a force driving it toward the right, as seen in the third frame. The final frame shows that the domain wall has successfully snapped onto the correct notch without affecting the other domain walls.

In essence, this method causes the domain wall oscillate around its initial position until it reaches a notch.

\section{Conclusion}
We derived an equation of motion for domain walls under applied normal strain. Comparison with mumax$^+$ simulations shows good agreement, except in cases where shear strain is present, such as for Rayleigh waves. In these cases, shear strain introduces an additional force that is not captured by the derivation and modifies the domain wall profile, leading to a breakdown of the Walker ansatz. Nevertheless, the model remains valid as long as the shear component is smaller than the normal component, which is typically the case for Rayleigh waves. Further work could investigate the role of shear strain in more detail.

We further demonstrated that a standing surface acoustic wave, formed by two counter-propagating Rayleigh waves, can be used as an error-correction mechanism in racetrack memory.

\section{Acknowledgments}
We thank Dirk Grundler for discussions in the early stage of the project.
D.D.G. and B.V.W. acknowledges financial support from the SHAPEme project (EOS ID 400077525) from the FWO and F.R.S.-FNRS under the Excellence of Science (EOS) program and the Ghent University special research fund  (bof/baf/4y/2024/01/288 ).

\bibliography{bibliography}

\begin{appendices}
\section{Integration of the LLG equation}\label{app:math}
We continue the derivation from \cref{EOM} to \cref{EOM DW} by multiplying by $\partial_x \mathbf{n}$ and integrating over the volume of the magnet
\begin{align}
\begin{split}
    -\frac{\mu_0M_\text{s} a^2}{8\gamma A_{0}}\int\mathbf{\ddot{n}}\cdot\partial_x\mathbf{n}\,\text{d}x =  &\gamma\int\mathbf{f_n}\cdot\partial_x\mathbf{n}\,\text{d}x\\
    &-\alpha\int\mathbf{\dot{n}}\cdot\partial_x\mathbf{n}\,\text{d}x, \label{EOM int}
\end{split}
\end{align}
where $t_y$ and $t_z$ are the width and thickness of the magnet, respectively. To proceed, we define $\xi = \frac{x-\chi(t)}{\Delta}$ and rewrite the derivatives in the collective coordinate frame.
\begin{align}
    \partial_x \mathbf{n} &= \frac{1}{\Delta} \frac{\partial \mathbf{n}_0}{\partial \xi}\\
    \mathbf{\dot{n}} &= \left(-\xi\frac{\dot{\Delta}}{\Delta}-\frac{\dot{\chi}(t)}{\Delta}\right)\frac{\partial\mathbf{n}_0}{\partial\xi}\\
    \begin{split}
    \mathbf{\ddot{n}} &= \left(-\xi\frac{\ddot{\Delta}}{\Delta}-\frac{\ddot{\chi}}{\Delta}+\xi\frac{2\dot{\Delta}^2}{\Delta^2}+\frac{2\dot{\Delta}\dot{\chi}}{\Delta^2}\right)\frac{\partial\mathbf{n}_0}{\partial\xi} \\
    &\hphantom{=}+ \left(-\xi\frac{\dot{\Delta}}{\Delta}-\frac{\dot{\chi}(t)}{\Delta}\right)^2\frac{\partial^2\mathbf{n}_0}{\partial\xi^2}.
    \end{split}
\end{align}
These expressions can now be used in \cref{EOM int}. The integrals can be evaluated by changing the integration variables to $\theta$ and integrating from $0$ to $\pi$, while retaining only terms up to first order in strain. Doing this for $\int\ddot{\mathbf{n}}\cdot\partial_x\mathbf{n}\,\text{d}x$ yields five different integrals given by
\begin{align}
    \left(-\frac{\ddot{\chi}}{\Delta}+\frac{2\dot{\Delta}\dot{\chi}}{\Delta^2}\right)\int \left(\frac{\partial\mathbf{n}_0}{\partial\xi}\right)^2 d\xi &= -\frac{2\ddot{\chi}}{\Delta^2}+\frac{4\dot{\Delta}\dot{\chi}}{\Delta^3}\\
    \left(-\frac{\ddot{\Delta}}{\Delta^2}+\frac{2\dot{\Delta}}{\Delta^2}\right)\int \xi \left(\frac{\partial\mathbf{n}_0}{\partial \xi}\right)^2 d\xi &= 0\\
    \frac{\dot{\Delta}^2}{\Delta^2}\int\xi^2\frac{\partial^2\mathbf{n}_0}{\partial\xi^2}\cdot\frac{\partial\mathbf{n}_0}{\partial\xi}d\xi &= 0\\
    \frac{2\dot{\Delta}\dot{\chi}}{\Delta^2}\int\xi\frac{\partial^2\mathbf{n}_0}{\partial\xi^2}\cdot\frac{\partial\mathbf{n}_0}{\partial\xi}d\xi &= -\frac{2\dot{\Delta}\dot{\chi}}{\Delta^3}\\
    \frac{\dot{\chi}^2}{\Delta^2}\int\frac{\partial^2\mathbf{n}_0}{\partial\xi^2}\cdot\frac{\partial\mathbf{n}_0}{\partial\xi}d\xi &= 0.
\end{align}
The next set of integrals are determined by $\int\dot{\mathbf{n}}\cdot\partial_x\mathbf{n}\,\text{d}x$ and result in two integrals.
\begin{align}
    -\frac{\dot{\Delta}}{\Delta}\int\xi\left(\frac{\partial\mathbf{n}_0}{\partial\xi}\right)^2 d\xi &= 0\\
    -\frac{\dot{\chi}(t)}{\Delta}\int\left(\frac{\partial\mathbf{n}_0}{\partial\xi}\right)^2 d\xi &= -\frac{2\dot{\chi}(t)}{\Delta^2}.
\end{align}
The final set of integrals are given by $\int\mathbf{f_n}\cdot\partial_x \mathbf{n}\,\text{d}x$. Both the integrals related to the inhomogeneous exchange $A_{11}$ and anisotropy $K$ result in $0$, while the strain can be evaluated to be
\begin{align}
\begin{split}
    \frac{-2B_1}{\Delta\mu_0M_\text{s}}&\int \left[\sum_{x,y,z}\varepsilon_{ii}n_{0i}\mathbf{e}_i\right]\cdot\partial_\xi\mathbf{n}_0\,\text{d}x\\
    &= \frac{-2B_1}{\mu_0M_\text{s}}\int_0^\pi (\varepsilon_{xx}-\varepsilon_{zz})\sin(\theta)\cos(\theta)\,\text{d}\theta.
\end{split}
\end{align}
 We can calculate this integral by using partial integration and only keeping terms up to first order in strain. This gives
\begin{align}
\begin{split}
    \frac{B_1}{\mu_0M_\text{s}}&\int_0^\pi \sin^2(\theta)\,\text{d}_\xi(\varepsilon_{xx}-\varepsilon_{zz})\frac{\text{d}\xi}{\text{d}\theta}\,\text{d}\theta\\
    &= \frac{B_1\Delta}{\mu_0M_\text{s}}\int_0^\pi \sin(\theta)\,\text{d}_\xi(\varepsilon_{xx}-\varepsilon_{zz})\,\text{d}\theta
\end{split}\\
    &= \frac{2B_1\Delta}{\mu_0M_\text{s}}\frac{\text{d}(\varepsilon_{xx}-\varepsilon_{zz})}{\text{d}x}.
\end{align}
Putting all of this in \cref{EOM int} results in the following equation of motion for a domain wall under the influence of normal strain
\begin{align}
\begin{split}
    \frac{\mu_0^2M_\text{s}^2 a^2}{8\gamma^2 A_{0}}\left(\frac{\ddot{\chi}}{\Delta}-\frac{\dot{\Delta}\dot{\chi}}{\Delta^2}\right)
    =  &B_1\Delta\frac{\text{d}(\varepsilon_{xx}-\varepsilon_{zz})}{\text{d}x}\\
    &+\alpha\mu_0M_\text{s}\frac{\dot{\chi}(t)}{\gamma\Delta}.
\end{split}
\end{align}
We can further evaluate this by deriving an expression for $\dot{\Delta}$.
\begin{align}
    \dot{\Delta} &= - \frac{\Delta B_1(\dot{\varepsilon}_{xx} - \dot{\varepsilon}_{zz})}{2(K+B_1(\varepsilon_{xx}-\varepsilon_{zz}))}-\frac{\Delta\dot{\chi}\ddot{\chi}}{c^2-\dot{\chi}^2}.
\end{align}
This provides the final expression for the equation of motion.
\begin{align}
    \frac{m_{\text{eff}}}{1-\frac{\dot{\chi}^2}{c^2}}\ddot{\chi} &= F - \left(\Gamma_\alpha + \Gamma_\Delta\right)\dot{\chi},
\end{align}
where we defined the effective mass, the force, and the damping parameters as
\begin{align}
    m_{\text{eff}} &= -\frac{\mu_0^2M_s^2 a^2}{8\gamma^2 A_0 \Delta}t_yt_z\\
    F &= -B_1\Delta\frac{\text{d}(\varepsilon_{xx}-\varepsilon_{zz})}{\text{d}x}t_yt_z\\
    \Gamma_\alpha &= \frac{\alpha\mu_0M_\text{s}}{\gamma\Delta}t_yt_z\\
    \Gamma_\Delta &= m_{\text{eff}}\frac{B_1(\dot{\varepsilon}_{xx} - \dot{\varepsilon}_{zz})}{2K_{\text{eff}}}.
\end{align}

\section{Shear Strain}\label{app:shear}
\begin{figure}
    \centering
    \includegraphics[width=\columnwidth]{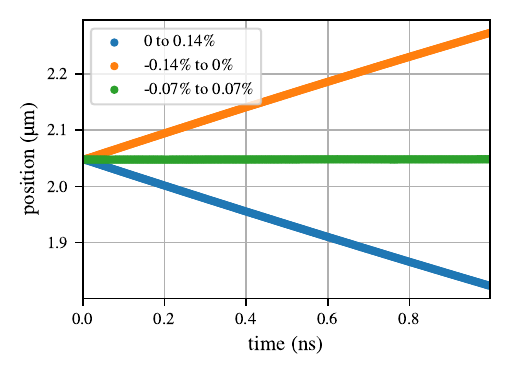}
    \caption{Position of a domain wall starting at the center of the simulation grid as a function of time under an applied shear strain gradient. The blue data is a gradient which start from \SI{0}{\%} strain at $x=\SI{0}{\micro\meter}$ and ends at \SI{0.14}{\%} at $x=\SI{4.096}{\micro\meter}$, while the orange data start from \SI{-0.14}{\%} and ends at \SI{0}{\%}. The green data corresponds to a gradient starting from \SI{-0.07}{\%} and ends at \SI{0.07}{\%}.}
    \label{fig:shear}
\end{figure}

As shown in \cref{fig:gradient_pos} the shear strain seems to force the domain wall towards lower strain values. However, \cref{fig:shear} shows three shear strain simulations withe \ch{NiO} parameters with different strain gradients, where the domain wall is always released from the center of the magnet. We can notice that in the simulation in which the strain goes from \SI{0}{\%} towards \SI{0.14}{\%} the domain wall moves towards lower strain values, while it moves towards higher strain values in the simulation where the strain goes from \SI{-0.14}{\%} towards \SI{0}{\%}. Finally we can see no movement when the domain in the simulation where the strain goes from \SI{-0.07}{\%} towards \SI{0.7}{\%}. From these simulations we can conclude that shear strain will force the domain wall towards $\varepsilon_{xz}=0$.

\section{Domain Wall Profile}\label{app:walker}
\begin{figure}
    \centering
    \includegraphics[width=\columnwidth]{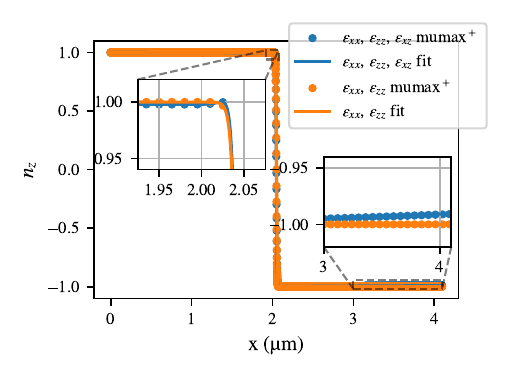}
    \caption{$z$-component of the Néel vector as a function of position. The dots are mumax$^+$ simulation data, while the lines are fits of the variable $x$ on that data using \cref{profile}. The blue data corresponds to a domain wall under an applied $\varepsilon_{xx}$, $\varepsilon_{zz}$ and $\varepsilon_{xz}$ strain, while the orange corresponds to a domain wall under an applied $\varepsilon_{xx}$ and $\varepsilon_{zz}$ strain. All applied strains have a gradient of $\SI{0.14}{\%}/\SI{4.096}{\micro\meter}$ at start at \SI{0}{\%} at $x=\SI{0}{\micro\meter}$.}
    \label{fig:profile}
\end{figure}

To find a profile for the domain wall under general shear and norm strain we start from the static energy density as a function of $\theta$ by using \cref{n0}
\begin{align}
    \begin{split}
        E = &2A_{11}(\partial_x\theta)^2 - 2K\cos^2(\theta)\\ 
    &+ 2B_1\left(\varepsilon_{xx}\sin^2(\theta)+\varepsilon_{zz}\cos^2(\theta)\right)\\
    &+ 2B_2\varepsilon_{xz}\left(\cos^2(\theta)-\sin^2(\theta)\right).
    \end{split}
\end{align}
The Euler-Lagrange equation then provides us with
\begin{align}
\begin{split}
    &A_{11}\partial_x^2\theta - (K+B_1(\varepsilon_{xx} - \varepsilon_{zz}))\cos(\theta)\sin(\theta)\\
    &+ B_2\varepsilon_{xz}(\cos^2(\theta) - \sin^2(\theta)) = 0.
\end{split}
\end{align}
This equation can be solved for $\theta$

\begin{align}
    &A_{11}\partial_x^2\theta = \frac{K_\text{eff}}{2}\sin(2\theta)-B_2\varepsilon_{xz}\cos(2\theta)\\
    &\frac{A_{11}}{2}\left(\partial_x\theta\right)^2 = \frac{K_\text{eff}}{2}\sin^2(\theta)-B_2\varepsilon_{xz}\cos(\theta)\sin(\theta)\\
    &x = \int\frac{d\theta}{\sqrt{\frac{K_\text{eff}}{A_{11}}(\sin^2(\theta)-\frac{B_2\varepsilon_{xz}}{K_\text{eff}}\cos(\theta)\sin(\theta))}}\\
    &x = \int\frac{\Delta d\theta}{\sqrt{\sin^2(\theta)-2\frac{B_2\varepsilon_{xz}}{K_\text{eff}}\cos(\theta)\sin(\theta)}}.
\end{align}
Here we have defined $K_\text{eff}=K+B_1(\varepsilon_{xx}-\varepsilon_{zz})$ and $\Delta = \sqrt{A_{11}/K_\text{eff}}$. We know that $B_2\varepsilon_{xz}/K_\text{eff}$ will be small and we can do a reverse Taylor expansion such that
\begin{align}
    x &= \int\frac{\Delta d\theta}{\sqrt{\sin^2\left(\theta+\frac{B_2\varepsilon_{xz}}{K_\text{eff}}\right)}}\\
    &= \Delta\ln\left(\tan\left(\theta/2-\frac{B_2\varepsilon_{xz}}{2K_\text{eff}}\right)\right)
\end{align}
As a final result for $\theta$ we have
\begin{align}\label{profile}
    \theta &= 2\arctan\left(e^{\frac{x}{\Delta}}\right) + \frac{B_2\varepsilon_{xz}}{K_\text{eff}}.
\end{align}
This equation reduces to the Walker ansatz when there is no shear strain. In \cref{fig:profile} this equation is fitted with variable $x$ to a domain wall where no shear strain is present and one where shear strain is present. It can be seen that this profile is a good fit and that the Walker ansatz can describe the domain wall when no shear strain is present.

\end{appendices}
\end{document}